# A MACRO LEVEL SCIENTOMETRIC ANALYSIS OF WORLD TRIBOLOGY RESEARCH OUTPUT (1998 - 2012)

Bakthavachalam Elango[1*], Periyaswamy Rajendran[2] and Lutz Bornmann[3]

[1]IFET College of Engineering,
Villupuram,
India.
Email: elangokb@yahoo.com

[2]SRM University,
Chennai,
India.
Email: librarian@srmuniv.ac.in

[3]Division for Science and Innovation Studies,
Administrative Headquarters of the Max Planck Society,
Hofgartenstr. 8,
80539 Munich, Germany.
Email: bornmann@gv.mpg.de

*Corresponding Author

# ABSTRACT

**Aim**

The aim of this study is to compare the country output and citation impact as well as to assess the level of interdisciplinarity in the field of tribology research across the period 1998 – 2012.

**Data & Methods**

Bibliographic records related to tribology research were extracted from SCOPUS and Web of Science databases for the period of 15 years from 1998 to 2012. Macro-level scientometric indicators such as growth rate, share of international collaborative papers, citations per paper, and share of non-cited papers were employed. Further, the Gini coefficient and Simpson Index of Diversity were used. Two new relative indicators – Relative International Collaboration Rate (RICR) and Relative Growth Index (RGI) – are proposed in this study. The performance of top countries contributing more than 1000 papers across the study period was discussed. Contributions and share of continents and countries by income groups were examined. Further research contributions and citation impact of selected country groups such as the Developing Eight Countries (D8), the Association of Southeast Asian Nations (ASEAN), the Union of South American Nations (UNASUR) and the Emerging and Growth-Leading Economies (EAGLEs) countries were analyzed.

**Results**

High levels of interdisciplinarity exist in tribology research. Inequality of distribution between countries is highest for number of publications and citations. Asia outperforms the other world regions and China contributes most of the papers (25%), while the United States receives most of the citations (22%). 84% of total output was contributed by the Asiatic region, Western Europe and North America together. Publications from these three world regions received 88% of total citations. Around 50% of global research output was contributed by China, the United States and Japan.

**Conclusion**

Tribology research output has increased drastically over the 15 year period by a factor of three, from 951 in 1998 to 2773 in 2012. The number of countries engaged in tribology research also grew from 55 in 1998 to 85 in 2012.

## INTRODUCTION

The word "tribology" was coined by Jost [1] in a report in 1965 as a composition of two Greek words *tribos* and *logos*. Tribology is defined as the science and engineering of surface phenomena such as friction, wear, lubrication, adhesion, surface fatigue, erosion, etc [2]. It is multidisciplinary in nature, and includes mechanical engineering (especially machine elements such as journal and roller bearings and gears), materials science, with research into wear resistance, surface technology with surface topography analysis and coatings, and the chemistry of lubricants and additives [3]. Tribological applications include improving car engines, hip joints and cosmetics, shrinking devices to micrometer and nanometer scales, and expanding the range of temperatures, speeds, and chemical environments where devices operate [4]. Apart from engineering applications, tribology can also be applied to products such as hair conditioners, lipsticks, powders, etc [5]. The relatively younger subdisciplines of tribology are: nanotribology (tribological phenomena occurring at sub-micron or smaller scales), biotribology (the tribology of the human body and other organisms), green tribology (science and technology of the tribological aspects of ecological balance and of environmental and biological impacts), and tribochemistry (the interaction of lubricants and lubricant additives with surfaces under tribological stress).

| Table 1. Recent macro level scientometric studies | | |
|---|---|---|
| Author(s) and year | Research Area | Geographical Area |
| Patra & Chand [13] | Library and Information Science | SAARC (South Asian Association for Regional Cooperation) and ASEAN (Association of Southeast Asian Nations) |
| Karpagam, Gopalakrishnan & Ramesh Babu [14] | Nanotechnology | G15 (Group of 15)* |
| Leta, Thijs & Glanzel [15] | Science | Brazil and Latin America |
| Sombatsompap, et al [16] | Energy and Fuel | ASEAN (Association of Southeast Asian Nations) |
| Borsi & Schubert [17] | Agricultural and Food Science | Europe |
| Wiysonge C S, et al [18] | Childhood Immunization | Africa |
| Soterades, et al [19] | Biomedical | World regions |
| Clarke, et al [20] | Public Health | Europe |
| Tan, Goudarzlou & Chakrabarty [21] | Service Research | Asia |
| Asplund, Eriksson & Persson [22] | Human Stroke | World wide |
| Chinchilla-Rodriguez, et al [23] | Medical research | Latin America and Caribbean |
| Plotnikova & Rake [24] | Pharmaceutical research | Worldwide |
| *composed of countries from Latin America, Asia and Africa. Comprises 17 countries although the name has not changed. Algeria, Argentina, Brazil, Chile, Egypt, India, Indonesia, Iran, Jamaica, Kenya, Malaysia, Mexico, Nigeria, Senegal, Sri Lanka, Venezuela and Zimbabwe. | | |

According to van Raan [6], bibliometric methods have been used in many disciplines of science and engineering to measure scientific progress. Scientometric indicators are useful to help scientists and decision makers to obtain valuable information [7]. There are three levels of bibliometric studies: macro (countries, scientific disciplines), meso (research centers, university departments, scientific subdisciplines) & micro (single papers, individual researchers) [8 – 10]. Macro-indicators, especially national science indicators are standard tools in bibliometrics and provide a comprehensive picture of national research output in scientific fields [11]. Bibliometric analyses performed at the macro-level (e.g. countries) yield at best general assessments of fields as a whole, for instance, the quality of a country's performance in physics, chemistry, psychology or immunology [12]. Macro level scientometric studies have been carried out in the past in various research fields. A selection is listed in Table 1.

Recently, Elango, Rajendran and Manickaraj [25] analyzed the tribology research output in BRIC (Brazil, Russia, India and China) countries, Elango, Rajendran and Bornmann [26] examined the global nanotribology research output, and Rajendran, Elango and Manickaraj [27] analyzed India's contribution to world tribology research. As a final step in analyzing tribology research, publication output of countries and regions, and degree of interdisciplinarity are analyzed in this study.

**OBJECTIVES**

The objective of this paper is to analyze scientific productivity and its impact in the field of tribology research as reflected in SCOPUS (Elsevier) and Web of Science (WoS, Thomson Reuters) during 1998 – 2012 using macro-level indicators by world region, level of income and various country groups. Further, the interdisciplinarity level of tribology research is evaluated with the Simpson Index of Diversity and Gini coefficient is used to examine the distribution of publications and citations among the contributing countries (equality vs. inequality). Two new relative indicators – Relative International Collaboration Rate (RICR) and Relative Growth Index (RGI) – are proposed in this study.

**MATERIALS AND METHODOLOGY**

**Data Set**
SCOPUS and WoS were used to retrieve the bibliographic records related to tribology research for the period of 15 years from 1998 to 2012. The following keywords were used in the combined fields of title, abstract and keywords: *\*tribolog\* OR "tribosyst\*" OR "tribo-syst\*" OR "tribo-chem\*" OR "tribochem\*" OR "tribotechn\*" OR "tribo-physi\*" OR "tribophysi\*"*. The search in SCOPUS was carried out on 19.12.2013 and refined to restrict the literature to articles, conference papers and reviews [28]. The corresponding search in WoS was performed in March 2014. The WoS data was taken from an in-house database belonging to the Max Planck Society (Munich, Germany). This database was established and is maintained by the Max Planck Digital Library (MPDL, Munich, Germany). The WoS data was used solely to produce the co-authorship networks, which are described in the following.

Self-citations have been included in the analyses, because self-citations are seen as an essential part of the scientific communication process [29, 30]. After removal of duplicate records, 27952 SCOPUS and 15729 WoS records were considered for the present study.

**Data Cleaning**
The procedure to count the author's country of origin is provided in table 2. The fractional counting method was applied to give credit to all the contributing countries [31, 32].

| Table 2. Detailed data cleaning procedure (examples) | | |
|---|---|---|
| Author | Affiliation | Remarks |
| Maalekian, M. | Institute for Materials Science and Welding, Graz University of Technology, Graz, **Austria**, Department of Materials Engineering, University of British Columbia, | Only primary affiliation is considered and others discarded. |

|  | Vancouver, Canada |  |
|---|---|---|
| Lovell, M.R. | Department of Industrial Engineering, University of Wisconsin-Milwaukee, Milwaukee, WI 53201, **India** | Country of affiliation is corrected. |
| Sasaki, S. | National Institute of Advanced Industrial Science and Technology (AIST), Tsukuba, Ibaraki 305-8564, Japan, **American Ceramic Society, United States** | Author's professional associations are discarded. |
| Nakamura, E. | Toyota Motor Corporation | Internet is used to confirm the country of affiliation where it is not available. |

**Indicators, coefficients and tools which have been employed**

**Growth Rate**
Compound Annual Growth Rate (CAGR) is used to give an indication of yearly growth [33]:

$$CAGR = \frac{End\ Value}{Beginning\ Value}^{\frac{1}{n-1}} - 1$$

**Relative Growth Index**
Relative Growth Index (RGI) is obtained by dividing the growth rate of a specific country during a period by a corresponding growth rate of all countries during the same period. RGI = 1 indicates that a country's growth rate is equal to the global average; RGI > 1 indicates that a country's growth rate is greater than the world average and RGI < 1 indicates a lower growth rate.

**Share of International Collaborative Papers**
The Share of International Collaborative Papers (SICP) measures the internationally co-authored publications in the national total as well as the strength of co-publication links between countries [34].

**Relative International Collaboration Rate**
The Relative International Collaboration Rate (RICR) is obtained by dividing the percentage of international collaborative papers of a country by the percentage of international collaborative papers of all countries. It is a simplified version of the International Collaboration Index suggested by Carg and Padhi [35]. RICR = 1 indicates that a country's international collaboration rate is equal to the global rate; RICR > 1 (or RICR < 1) indicates that a country's international collaboration rate is greater (or lower) than the world average.

**Citation per paper**
Citation per paper (CPP) is obtained by dividing the total number of citations by the total number of papers.

**Non-Citation Relative Rate**
The Non-Citation Relative Rate (NCRR) is the quotient of the percentage of a country's non-cited papers and of all the countries. NCRR = 1 indicates that a country's uncitedness is equal to the world average; NCRR > 1 (NCRR < 1) indicates that a country's uncitedness is greater (lower) than the world average. NCRR = 0 indicates that a country's uncitedness is 0.

**International Collaboration Network**
Pajek (http://pajek.imfm.si/doku.php) is used to produce a co-authorship network of the most productive countries in tribology research.

**Gini Coefficient**
The Gini coefficient is commonly used as a measure of inequality of income or wealth. In this paper, we examine the degree of equality of publications and citations between countries. The highest possible Gini coefficient is 1. This means that one country gets all the publications credit. A coefficient of 0 means that publications and citations are equally distributed among the countries.

**Simpson Index of Diversity**
The multidisciplinary character of tribology can be measured on the base of SCOPUS subject areas [36]. Scopus classifies journal titles into 27 major subject areas, where a journal may belong to more than one subject area. The multidisciplinary nature of the tribology research can be assessed by the distribution of the papers across different subject areas. We use Simpson Index of Diversity to characterize this:

$$SID = 1 - \frac{\sum n\,(n-1)}{N\,(N-1)}$$

Where n is the number of papers attributed to the i[th] subject area and N is the total number of papers attributed to all subject areas. The value ranges between 0 and 1; the greater the SID, the greater the sample diversity.

**DATA ANALYSIS**

Table 3 provides a general overview over the tribology research output for the period of 15 years from 1998 to 2012. 97.5% of the tribology research papers have country affiliation information for the authors (in SCOPUS). 7.94% annual publication growth was observed over the period. The Gini indices show that a very small number of countries produced most of the publications and received most of the citations.

| Table 3. Summary of tribology research output during 1998 – 2012 | |
|---|---:|
| **SCOPUS based bibliographic records** | |
| Number of Papers | 27952 |
| CAGR | 7.94% |
| Countries involved | 108 |
| Information about country of origin of authors available | 27252 (97.5%) |
| International collaborative papers (%) | 3789 (13.9%) |
| Citations received by all papers from time of publication to 19.12.2013 | 238563 |
| Cited papers (%) | 20684 (74%) |
| CPP | 8.53 |
| Citation per paper per year (CPPY) | 1.45 |
| Gini Index for countries against publications | 0.844 |
| Gini Index for countries against citations | 0.852 |
| **WOS based bibliographic records** | |
| Number of Papers | 15729 |

General characteristics of tribology research from 1998 to 2012 are presented in Table 4: yearly output, CPP, and share of cited papers. A threefold increase was observed over the study period, (from 951 in 1998 to 2773 in 2012). The highest number of papers was published in the year 2011, with 3645 and the lowest in 1999 with 946.

| Table 4. Yearly output and citation impact 1998 – 2012 ||||  |
| Year | TP | TC | CPP | %cited |
|---|---|---|---|---|
| 1998 | 951 | 12580 | 13.23 | 68.24 |
| 1999 | 946 | 16026 | 16.94 | 81.18 |
| 2000 | 1017 | 16796 | 16.52 | 76.50 |
| 2001 | 1087 | 17190 | 15.81 | 75.53 |
| 2002 | 1144 | 13735 | 12.01 | 73.16 |
| 2003 | 1197 | 17642 | 14.74 | 79.78 |
| 2004 | 1467 | 19031 | 12.97 | 77.10 |
| 2005 | 1466 | 18627 | 12.71 | 79.81 |
| 2006 | 1502 | 17426 | 11.60 | 78.83 |
| 2007 | 1365 | 15082 | 11.05 | 80.88 |
| 2008 | 2223 | 16857 | 7.58 | 78.81 |
| 2009 | 3574 | 24236 | 6.78 | 78.76 |
| 2010 | 3595 | 17579 | 4.89 | 76.72 |
| 2011 | 3645 | 11484 | 3.15 | 69.79 |
| 2012 | 2773 | 4272 | 1.54 | 51.14 |
| TP = Total Papers, TC = Total Citations, CPP = Citation per paper ||||  |

According to the Science and Engineering Indicators [37], collaboration across national boundaries is generally increasing. Tribology research is no exception. The share of international collaborative papers has increased from 10% in the first five year period to almost 15% in the last five year period (see table 5). 14% of all papers were published with international collaboration across the study period.

| Table 5. Publication output in five year block periods ||||
| Block Period | ICP | TP | % |
|---|---|---|---|
| 1998-2002 | 481 | 4794 | 10.03 |
| 2003-2007 | 993 | 6800 | 14.60 |
| 2008-2012 | 2315 | 15658 | 14.78 |
| ICP = International Collaborative Papers, TP = Total Papers ||||

Table 6 shows the comparison between international collaboration and national output. International collaborative papers received higher CPP than national output. This result is in agreement with the result of many other studies [see e.g. 38].

| Table 6. International vs. national output ||||  |
| Type of Collaboration | TP | TC | CPP | %cited |
|---|---|---|---|---|
| International Collaboration | 3789 | 44340 | 11.70 | 85.14 |
| Single Country | 23463 | 192598 | 8.21 | 73.38 |
| TP = Total Papers, TC = Total Citations | | | | |

Classification of countries by world regions is adopted from SCImago (www.scimagojr.com). Table 7 presents the contribution and share of the world regions. Almost 46% of world tribology research output was contributed by authors located in the Asiatic region followed by Western Europe, North America and Eastern Europe. Africa had the lowest contribution among the world regions even though the number of contributing countries is higher than the Middle East. Publications from North America received the highest citations per paper (14.5) followed by Western Europe with 11.8 and the Middle East with 9.6. Among the continents, North America and Pacific Region received the lowest Gini Index value while Asia received the highest. A low Gini Index value represents equally contributing countries within regions and index value near 1 represent a concentration of publications in certain countries. Note that North America and the Pacific Region each consist of only two countries (see table 7).

**Table 7.** Contribution by world regions

| Region | # Countries | TP | World Share | CPP | Gini Index for Publications | Leading Country |
|---|---:|---:|---:|---:|---:|---|
| Asiatic Region | 18 | 12424.16 | 45.59 | 6.10 | 0.83 | China |
| Western Europe | 21 | 6576.22 | 24.13 | 11.77 | 0.64 | Germany |
| North America | 2 | 3939.46 | 14.46 | 14.51 | 0.39 | United States |
| Eastern Europe | 22 | 2465.84 | 9.05 | 4.48 | 0.67 | Russian Federation |
| Middle East | 14 | 996.78 | 3.66 | 9.59 | 0.74 | Turkey |
| Latin America | 12 | 482.77 | 1.77 | 6.64 | 0.74 | Brazil |
| Pacific Region | 2 | 240.20 | 0.88 | 8.92 | 0.43 | Australia |
| Africa | 17 | 126.07 | 0.46 | 4.95 | 0.71 | South Africa |
| TP = Total Papers, CPP = Citation per paper | | | | | | |

The classification of countries by income group was obtained from the World Bank (http://data.worldbank.org). The distribution of tribology contributions by income group is presented in table 8. It can be observed that there is a relationship between the income of a specific country and its research activity. Almost 95% of world publications are from countries of the high and upper middle income categories. As expected, publications from high and lower income countries have the highest CPP. The lower income countries profit from larger proportions of papers with international collaboration: Out of the eight countries in the lower income group, five published all their papers in international collaboration.

**Table 8.** Contribution by income groups

| Income Group | # Countries | TP | World Share | CPP |
|---|---:|---:|---:|---:|
| High Income | 46 | 16233.79 | 59.57 | 10.88 |
| Upper Middle Income | 34 | 9516.74 | 34.92 | 5.13 |
| Lower Middle Income | 20 | 1477.48 | 5.42 | 7.63 |
| Lower Income | 8 | 23.49 | 0.09 | 10.17 |
| TP = Total Papers, CPP = Citations Per Paper | | | | |

Table 9 provides information about the publication patterns of the top 7 countries in the dataset, which published more than 1000 papers over the study period. Except for China and India, five countries belong to the G7 group (USA, UK, France, Germany, Italy, Canada and Japan). This shows that the G7 nations are the leaders in tribology research. The seven countries in the table 9 together contributed 66.5% of the world output. Among the top countries, China contributed 25% of the total output, followed by the United States with 13% and Japan with 10.5%. Except for China and Japan, the authors of all the top countries have collaborated with authors from other countries above the world collaboration rate (RICR). India tops the list in the papers' growth rate with 19%. With RGIs greater than 1, only India's and China's growth rates were higher than the world average. Among the top seven countries, five belong to the high income group.

**Table 9.** Contributions of top countries (>1000 papers)

| Country | TP | World Share | #ICP | RICR | Growth Rate in % | RGI | Income Group |
|---|---|---|---|---|---|---|---|
| China | 6759.27 | 24.80 | 719 | 0.73 | 15 | 1.86 | Upper Middle |
| United States | 3524.08 | 12.93 | 1098 | 1.92 | 3 | 0.37 | High Income |
| Japan | 2870.33 | 10.53 | 431 | 1 | 2 | 0.28 | High Income |
| Germany | 1631.28 | 5.99 | 594 | 2.19 | 6 | 0.70 | High Income |
| United Kingdom | 1195.58 | 4.39 | 578 | 2.76 | 7 | 0.91 | High Income |
| India | 1073.88 | 3.94 | 201 | 1.23 | 19 | 2.44 | Lower Middle |
| France | 1070 | 3.93 | 522 | 2.79 | 5 | 0.67 | High Income |
| TP = Total Papers, ICP = International Collaborative Papers, RICR = Relative International Collaboration Rate, RGI = Relative Growth Index | | | | | | | |

The citation impact of the top countries is provided in table 10. The papers of these top countries received 66% of world citations. Among the top countries, contributions from the United States received 22% of world citations and contributions from India only 4%. However, India has the fourth highest CPP of 9.18. Contributions from the United Kingdom have the highest CPP of 15.17, and the lowest CPP applies to contributions from China and Japan. Except for China, the NCRR of all the top countries is lower than the world average of 1.

**Table 10.** Citation impact of top countries (>1000 papers)

| Country | TC | %TC | CPP | NCRR |
|---|---|---|---|---|
| China | 34351.88 | 14.50 | 5.08 | 1.19 |
| United States | 52160.6 | 22.01 | 14.80 | 0.29 |
| Japan | 15174.52 | 6.40 | 5.29 | 0.66 |
| Germany | 14112.97 | 5.96 | 8.65 | 0.28 |
| United Kingdom | 18134.37 | 7.65 | 15.17 | 0.11 |
| India | 9854.38 | 4.16 | 9.18 | 0.11 |
| France | 12976.65 | 5.48 | 12.13 | 0.10 |
| TC = Total Citations, CPP = Citation Per Paper, NCRR = Non Citation Relative Rate | | | | |

Table 11 presents the most productive UNASUR (Union of South American Nations) countries from 1998 to 2012. Among the countries, Brazil contributed more than 1% of world publication output during the study period, followed by Colombia and Argentina. Only Ecuador received a higher CPP value than the world average of 8.53 (see table 3). Except for Brazil and Colombia, the NCRR of all countries is lower than the world average of 1.

**Table 11.** Contribution and impact of UNASUR countries

| Country | TP | World Share | CPP | %ICP | NCRR |
|---|---|---|---|---|---|
| Brazil | 283.33 | 1.04 | 6.89 | 33.43 | 6.27 |
| Colombia | 42.67 | 0.16 | 6.17 | 60.94 | 1.14 |
| Argentina | 39.50 | 0.14 | 6.21 | 48.15 | 0.85 |
| Venezuela | 26.92 | 0.1 | 7.70 | 71 | 0.85 |
| Chile | 12.75 | 0.05 | 1.84 | 47.06 | 0.64 |
| Ecuador | 0.50 | 0.002 | 9.00 | 100.00 | 0.00 |
| Total | 405.67 | 1.49 | 6.65 | | |

TP = Total Papers, CPP = Citation Per Paper, ICP = International Collaborative Papers, NCRR = Non Citation Relative Rate

Table 12 shows the publication pattern of ASEAN (Association of Southeast Asian Nations) countries. Out of 10 ASEAN countries, only six countries engaged in the research field of tribology during the study period, and these six countries together contributed 1.7% of total world output. Singapore is the top ASEAN country with the highest publication share, and Thailand received the highest CPP. Among the ASEAN countries, Singapore and Thailand received a higher CPP than the world average of 8.53 (see table 3). Vietnam contributed all the papers with international collaboration. All ASEAN countries had higher shares of international collaborative papers than the world average. All countries received lower NCRR than the world average.

**Table 12.** Contribution and impact of ASEAN countries

| Country | TP | World Share | CPP | %ICP | RICR | NCRR |
|---|---|---|---|---|---|---|
| Singapore | 225.25 | 0.83 | 12.68 | 29.96 | 2.16 | 0.46 |
| Malaysia | 125.83 | 0.46 | 7.23 | 28.57 | 2.05 | 0.84 |
| Thailand | 64.27 | 0.24 | 14.20 | 33.33 | 2.40 | 0.35 |
| Colombia | 42.67 | 0.16 | 6.17 | 60.94 | 4.38 | 0.96 |
| Indonesia | 6 | 0.02 | 9.25 | 90.91 | 6.54 | 0.35 |
| Vietnam | 2.33 | 0.01 | 5.94 | 100.00 | 7.19 | 0.77 |

| Total | 466.35 | 1.71 | 10.74 | | | |

TP = Total Papers, CPP = Citation Per Paper, ICP = International Collaborative Papers, RICR = Relative International Collaboration Rate, NCRR = Non Citation Relative Rate

All the countries of D8 (Developing Eight) are engaged in tribology research (see table 13). Among the D8 countries, Turkey and Iran contributed more than 1% of total world output and all the D8 countries together contributed 4% of total world output. Publications from Bangladesh received the highest citations per paper (11.2) followed by Pakistan with 9.5 and Indonesia with 9.25. Indonesia produced most of its papers in international collaboration, as did Pakistan and Bangladesh. Among the D8 countries, Egypt, Bangladesh and Nigeria received a higher NCRR than the world average.

**Table 13.** Contribution and impact of D8 countries

| Country | TP | World Share | CPP | %ICP | RICR | NCRR |
|---|---|---|---|---|---|---|
| Turkey | 407.45 | 1.68 | 7.02 | 17.41 | 1.25 | 0.77 |
| Iran | 283.17 | 1.16 | 5.60 | 18.27 | 1.31 | 0.80 |
| Malaysia | 125.83 | 0.57 | 7.23 | 28.57 | 2.05 | 0.84 |
| Egypt | 78.50 | 0.37 | 4.46 | 30.11 | 2.17 | 1.12 |
| Bangladesh | 16.83 | 0.10 | 11.17 | 52.17 | 3.75 | 1.00 |
| Pakistan | 16 | 0.10 | 9.47 | 76.92 | 5.53 | 0.74 |
| Nigeria | 13.50 | 0.06 | 1.33 | 20.00 | 1.44 | 2.31 |
| Indonesia | 6 | 0.05 | 9.25 | 90.91 | 6.54 | 0.35 |
| Total | 947.28 | 3.48 | 6.46 | | | |

TP = Total papers, CPP = Citation Per Paper, ICP = International Collaborative Papers, RICR = Relative International Collaboration Rate, NCRR = Non Citation Relative Rate

EAGLEs (Emerging and Growth-Leading Economies) countries together contributed 38% of the tribology output, where China is the leader followed by India, Russian Federation and South Korea (see table 14). Apart from Egypt, Mexico and Indonesia, all D8 countries contributed more than 1% of the world's total output. The share of international collaborative papers for China and Taiwan is lower than the world average. Among the D8 countries, the share of non-cited papers for China, the Russian Federation, Egypt and Mexico is higher than the world average.

**Table 14.** Contribution and impact of EAGLEs countries

| Country | TP | World Share | CPP | %ICP | RICR | NCRR |
|---|---|---|---|---|---|---|
| China | 6759.25 | 24.80 | 5.08 | 10.08 | 0.73 | 1.19 |
| India | 1073.88 | 3.94 | 9.18 | 17.05 | 1.23 | 0.65 |
| Russian Federation | 680.03 | 2.50 | 3.73 | 23.23 | 1.67 | 1.84 |
| South Korea | 644.08 | 2.36 | 8.67 | 27.20 | 1.96 | 0.67 |
| Taiwan | 484.67 | 1.78 | 8.54 | 9.43 | 0.68 | 0.46 |
| Turkey | 407.45 | 1.50 | 7.02 | 17.41 | 1.25 | 0.77 |
| Brazil | 283.33 | 1.04 | 6.89 | 33.43 | 2.40 | 0.98 |
| Egypt | 78.50 | 0.29 | 4.46 | 30.11 | 2.17 | 1.12 |

| Mexico | 68 | 0.25 | 6.20 | 52.13 | 3.75 | 1.10 |
| Indonesia | 6 | 0.02 | 9.25 | 90.91 | 6.54 | 0.35 |
| Total | 10485.19 | 38.47 | 5.92 | | | |
| TP = Total Papers, CPP = Citation Per Paper, ICP = International Collaborative Papers, RICR = Relative International Collaboration Rate, NCRR = Non Citation Relative Rate ||||||| 

**International collaboration network**

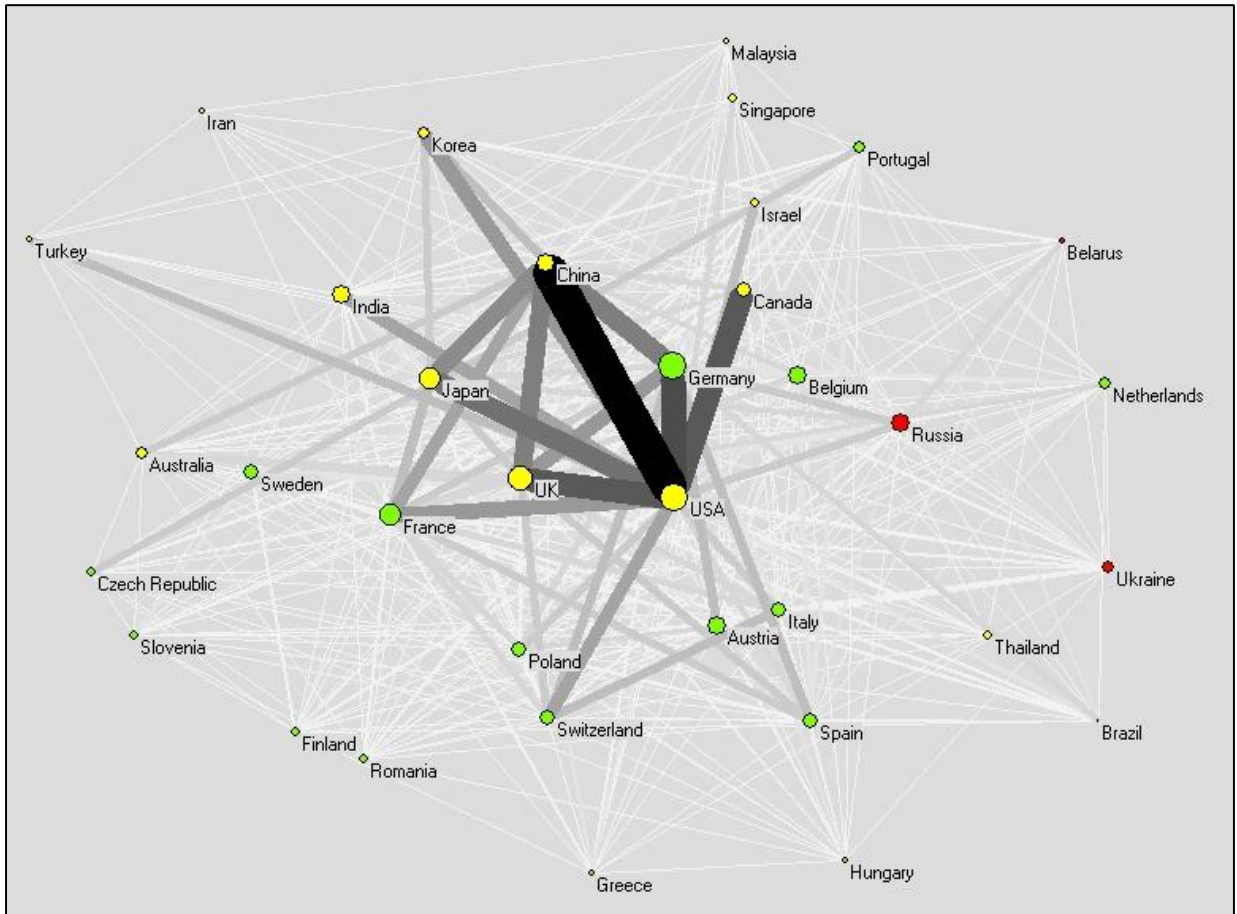

**Figure 1.** Co-authorship relations between countries engaging in tribology research

Figure 1 shows the co-authorship relations between countries with a degree (number of links in the network) of 12 or more. The size of the vertices is determined by the betweenness centrality of the countries. In order to find an optimal layout for the network, the spring embedder of Kamada and Kawai [39] is used. The community-finding algorithm of Blondel et al. [40] distinguishes three groups: (1) a yellow group with USA, UK, China, and Japan at the core of this network, and some other countries at the periphery (e.g. India, Korea, and Iran); (2) a green group with only EU countries (with the exception of UK); and (3) a small red group consisting of Russia and the Ukraine.

**Measurement of interdisciplinarity**

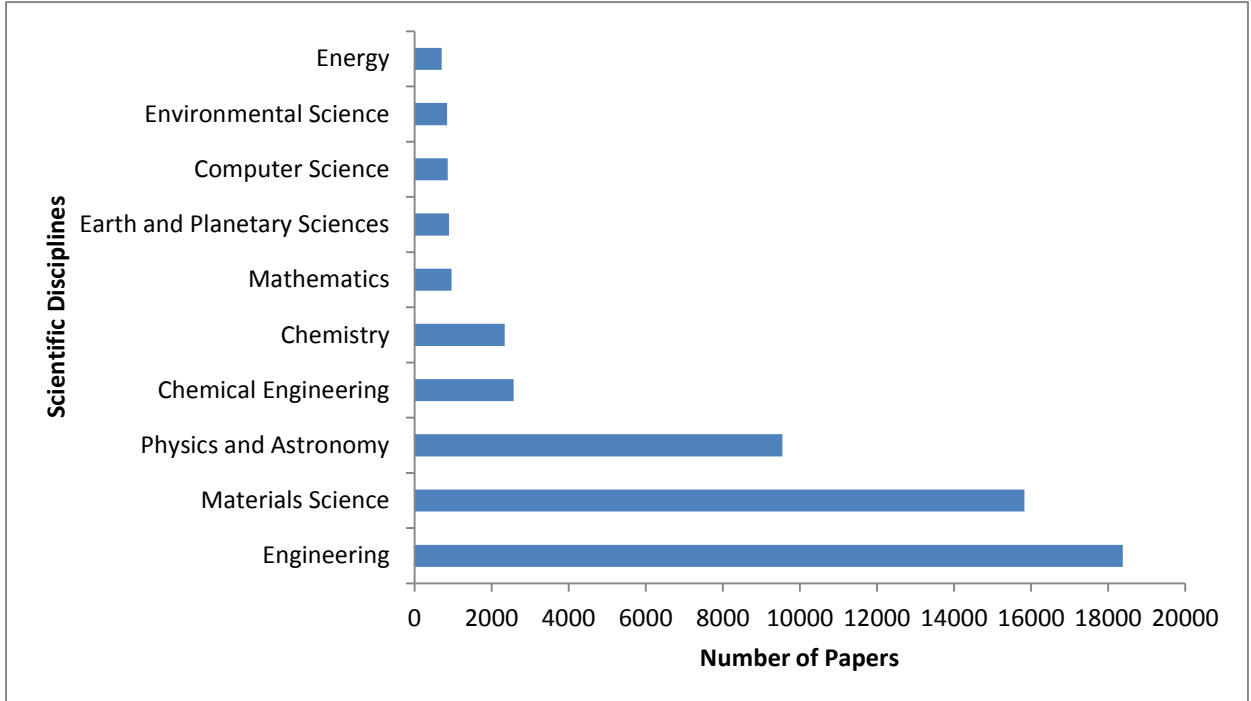

**Figure 2**. SCOPUS Subject areas attributed by tribology research papers

In this study, we use the Simpson Index of Diversity based on the number of SCOPUS subject areas to measure the level of interdisciplinarity in tribology research [41]. This index is commonly used for calculating biodiversity habitats in ecology. For example, the degree of interdisciplinarity has been assessed in the fields of Forestry [42] and Cardiovascular Systems [43]. The analyses of interdisciplinarity are based solely on those papers in the dataset of this study which are indexed under the main SCOPUS subject category - Physical Sciences. Tribology research belongs to pure engineering and nearly all papers have been categorized in this main category.

The value of the Simpson Index of Diversity is calculated as 0.75, which shows the high level of interdisciplinarity in tribology research. Figure 2 shows the different subject areas of the papers (through journals) in tribology research. It can be observed from Figure 2 that all the papers in the dataset have been attributed to either Engineering or Materials Science along with other subject areas.

**DISCUSSION& LIMITATIONS**

We examined the world tribology research output across a 15 years period. There were 108 countries involved in tribology research during this period. Tribology research work is dominated by the Asiatic region and high income countries. Similar results have been reported for related disciplines such as materials science [44]. The results based on the Gini Index show that a very small group of countries were responsible for most of the publications and received most of the citations in this field. There exists a high level of interdisciplinarity in the tribology papers. The share of international collaborative papers is 13.9% which is lower than for other research fields such as stem cell with 21% [45]. China contributed 25% of the world's total tribology research output during the study period, which is a higher share than in other research fields such as Global Positioning System [46], stem cell [47] and medicine [48]. In these fields, China's contribution was below 10%. Contributions by authors from North America had the highest impact and those from Eastern Europe the lowest.

There are two limitations of this study. Firstly, growth rates have been calculated using CAGR which is based entirely on the initial and final values. It takes no account of changes in-between. Secondly, the level of interdisciplinarity has been evaluated based on the subject classifications at journal level. Interdisciplinarity should actually be measured on the basis of individual papers.

**ACKNOWLEDGMENTS**

The Web of Science data used in this paper is from a bibliometrics database developed and maintained by the Max Planck Digital Library (MPDL, Munich) and derived from the Science Citation Index Expanded (SCI-E), Social Sciences Citation Index (SSCI), Arts and Humanities Citation Index (AHCI) prepared by Thomson Reuters (Scientific) Inc. (TR®), Philadelphia, Pennsylvania, USA: ©Copyright Thomson Reuters (Scientific) 2014.

The authors are grateful to Dr. Yu Tian, Professor at Tsinghua University, China for his constructive comments on improving the manuscript.

**REFERENCES**

1. Jost P (1966). *Lubrication (Tribology) – A Report on the Present Position and Industry's Needs*. Dept. of Education and Science (H.M. Stationary Office).
2. www.engineeringmaterials.org/tribology (accessed on 08.10.2013)
3. Mang T, Bobzin K and Bartel T (2011). Industrial Tribology : Tribosystems, Friction, Wear and Surface Engineering, Lubrication. Wiley-Wch, Weinheim.
4. http://www.grc.org/conferences.aspx?id=0000277(accessed on 08.10.2013)
5. http://www.jytra.com/blog/technical/what-is-industrial-tribology.html (accessed on 08.10.2013)
6. van Raan A F J (2005). For your citations only? Hot topics in bibliometric analysis measurement. *Interdisciplinary Research and Perspectives*, 3: 50 – 62.
7. Jin B & Rousseau R (2004). Evaluation of research performance and scientometrics in china. In H. F. Moed, W. Glänzel & U. Schmoch (Eds.), *Handbook of Quantitative Science and Technology Research* (pp. 497-514). Dordrecht, etc.: Kluwer Academic Publishers.
8. Vinkler P (1988). An attempt of surveying and classifying bibliometric indicators for scientometric purposes. *Scientometrics*, 13(5-6): 239 – 259.
9. Glänzel W & Moed H F (2002). Journal impact measures in bibliometric research. *Scientometrics*, 53(2): 171 – 193.
10. Fiala D (2013). Suborganizations of institutions in library and information science journals. *Information*, 4: 351 – 366.
11. Moed H F, Glänzel W & Schmoch U (2004). Editors' Introduction. In H. F. Moed, W. Glänzel & U. Schmoch (Eds.), *Handbook of Quantitative Science and Technology Research* (pp. 1-15). Dordrecht, etc.: Kluwer Academic Publishers.
12. van Raan A F J (2003). The use of bibliometric analysis in research performance assessment and monitoring of interdisciplinary scientific developments. *Technikfolgenabschätzung*, 1: 20-29
13. Patra S K & Chand P (2009). Library and information science research in SAARC and ASEAN countries as reflected through LISA. *Annals of Library and Information Studies*, 56(1): 41-51.
14. Karpagam R, Gopalakrishnan S & Ramesh Babu B (2011). Publication trend on nanotechnology among G15 countries: A bibliometric study. *COLLNET Journal of Scientometrics and Information Management*, 5(1): 61.
15. Leta J, Thijs B & Glänzel W (2013). A macro-level study of science in Brazil: Seven years later. *EncontrosBibli*, 18(36): 51-66.
16. Sombatsompop N, et al (2011). Research productivity and impact of ASEAN countries and universities in the field of energy and fuel. *Malaysian Journal of Library and Information Science*, 16(1): 35-46.
17. Borsi B & Schubert A (2011). Agrifood research in Europe: a global perspective. *Scientometrics*, 86: 133 – 154.
18. Wiysonge C S, et al (2013). A bibliometric analysis of childhood immunization research productivity in Africa since the onset of the Expanded Program on Immunization in 1974. *BMC Medicine*, 11:66. Available at: http://www.biomedcentral.com/1741-7015/11/66
19. Soteriades E S, et al (2005). Research contribution of different world regions in the top 50 biomedical journals (1995 – 2002). *FASEB Journal*, 20: 29-34.

20. Clarke A, et al (2007). A bibliometric overview of public health research in Europe. *European Journal of Public Health*, 17(Sup.1): 43-49.
21. Tan K C, Goudarzlou A & Chakrabarty A (2010). A bibliometric analysis of service research from ASIA. *Managing Service Quality*, 20(1): 89-101.
22. Asplund K, Eriksson M & Persson O (2012). Country comparisons of human stroke research since 2001: A bibliometric study. *Stroke*, 43: 830-837.
23. Chinchilla-Rodriguez Z, et al (2012). International collaboration in medical research in Latin America and the Caribbean (2003-2007). *Journal of American Society of Information Science and Technology*, 63(11): 2223-2238.
24. Plotnikova T & Rake B (2014). Collaboration in pharmaceutical research: exploration of country-level determinants. *Scientometrics*, 98(2): 1173-1202.
25. Elango B, Rajendran P & Manickaraj J (2013). Tribology research output in BRIC countries: A scientometric dimension. *Library Philosophy and Practice*, Paper 935.
26. Elango B, Rajendran P & Bornmann L (2013). Global nanotribology research output (1996 – 2010): A scientometric analysis. *PLOS ONE*, 8(12): e81094.
27. Rajendran P, Elango B & Manicakraj J (2014). Publication trends and citation impact of tribology research in India: A scientometric study. *Journal of Information Science Theory and Practice,* 2(1): 22-34.
28. Carg K C, et al (2010). Scientoemtric profile of genetics and heredity research in India. *Annals of Library and Information Studies*, 57(3): 196–206.
29. Glänzel W (2003). Bibliometrics as a research field: a course on theory and application of bibliometric indicators. Course Handouts.
30. Leta J, Thijs B & Glänzel W (2013). A macro-level study of science in Brazil: seven years later. *EncontrosBibli*, 18(36): 51 – 66.
31. Borsi B & Schubert A (2011). Agrifood research in Europe: a global perspective. *Scientometrics*, 86: 133 – 154.
32. Elango B, Rajendran P & Bornmann L (2013). Global nanotribology research output (1996 – 2010): A scientometric analysis. *PLOS ONE*, 8(12): e81094.
33. Choi D G, Lee H & Sung T K (2011). Research profiling for 'standardization and innovation'. *Scientometrics*, doi: 10/1007/s/11192-011-0344-7.
34. Glänzel W (2000). A bibliometric analysis of co-authorship patterns of Eleven East Central European Countries in the 90s. Proceedings of the Second Berlin Workshop on Scientometrics and Informetrics, Berlin.
35. Carg K C & Padhi P (2001). A study of collaboration in laser science and technology. *Scientometrics*, 51: 415-427.
36. Igami M & Saka A (2007). Capturing the evolving nature of science, the development of new scientific indicators and the mapping of science. OECD Science, Technology and Industry Working Papers, 2007/1. Available at: http://www.oecd-ilibrary.org
37. National Science Board (2012). *The overview of Science and Engineering Indicators.* Washington DC: National Science Foundation; Available at: http://www.nsf.gov/statistics/seind12/pdf/overview.pdf
38. Glänzel W (2001). National characteristics in international scientific co-authorship relations. *Scientometrics*, 51(1): 69-115.
39. Kamada T & Kawai S (1989). An algorithm for drawing general undirected graphs. *Information Processing Letters*, 31: 7-15.

40. Blondel et al. (2008). Fast unfolding of communities in large networks. *Journal of Statistical Mechanics-Theory and Experiment*. doi: 10.1088/1742-5468/2008/10/P10008.
41. Kalz M & Specht M (2013). Assessing the crossdisciplinary of technology-enhanced learning with science overlay maps and diversity measures. *British Journal of Educational Technology*, DOI: 10.1111/bjet.12092.
42. Bojović S, et al (2014). An overview of forestry journals in the period 2006–2010 as basis for ascertaining research trends. *Scientometrics*, 98(2): 1331-1346.
43. Leydesdorff L and Opthof T (2013). Citation analysis with medical subject Headings (MeSH) using the Web of Knowledge: A new routine. *Journal of the American Society for Information Science and Technology*, 64(5): 1076-1080.
44. Kademani B S, et al (2013). Publication trends in materials science: a global perspective. *Scientometrics*, 94: 1275 – 1295.
45. Luo J & Matthews K R W (2013). Globalization of stem cell science: An examination of current and past collaborative research networks. *PLOS ONE*, 8(9): e73598.
46. Wang H, et al (2013). A historical review and bibliometric analysis of GPS research from 1991 – 2010. *Scientometrics*, 95: 35-44.
47. Luo J & Matthews K R W (2013). Globalization of stem cell science: An examination of current and past collaborative research networks. *PLOS ONE*, 8(9): e73598.
48. Gupta B M & Bala A (2011). A scientometric analysis of Indian research output in medicine during 1999–2008. *Journal of Natural Science, Biology and Medicine*, 2(1): 87-100.